\documentclass[conference,10pt,letterpaper]{IEEEtran}
\usepackage{amsmath,cite,amsfonts,amssymb,psfrag,amsthm,paralist}
\usepackage{graphicx}
\usepackage{epstopdf}
\usepackage{rotating}
\usepackage{dsfont}
\usepackage{color}
\usepackage{tikz}
\usetikzlibrary{plotmarks}
\usepackage{pgfplots}
\usetikzlibrary{calc}
\usetikzlibrary{shapes,arrows}
\usetikzlibrary{decorations.markings}
\usetikzlibrary{positioning}
\pgfplotsset{compat=1.10}
\usetikzlibrary{calc}
\usetikzlibrary{shapes,arrows}
\usetikzlibrary{decorations.markings}

\usepackage[utf8]{inputenc}

\DeclareFontFamily{U}{mathx}{\hyphenchar\font45}
\DeclareFontShape{U}{mathx}{m}{n}{<-> mathx10}{}
\DeclareSymbolFont{mathx}{U}{mathx}{m}{n}
\DeclareMathAccent{\widebar}{0}{mathx}{"73}

\newcommand{\Enc}{\mathsf{Enc}}
\newcommand{\Dec}{\mathsf{Dec}}

\makeatletter
\newtheorem*{rep@theorem}{\rep@title}
\newcommand{\newreptheorem}[2]{%
	\newenvironment{rep#1}[1]{%
		\def\rep@title{\Cref{##1}}%
		\begin{rep@theorem}}%
		{\end{rep@theorem}}}
\newcommand*{\textlabel}[2]{%
	\edef\@currentlabel{#1}
	\phantomsection
	#1\label{#2}
}
\makeatother

\newtheorem{theorem}{Theorem}

\newtheorem{definition}{Definition}

\newtheorem{lemma}{Lemma}
\newtheorem{corollary}{Corollary}

\usepackage[free-standing-units]{siunitx}
\usepackage{circuitikz}
\usetikzlibrary[arrows.meta,backgrounds,bending]

\begin{document}
\title{Secure and Private Source Coding with Private Key and Decoder Side Information}
\IEEEoverridecommandlockouts

\author{%
	\IEEEauthorblockN{Onur G\"unl\"u\textsuperscript{1,2},
		Rafael F. Schaefer\textsuperscript{1}, Holger Boche\textsuperscript{3,4,5}, and
		H. Vincent Poor\textsuperscript{6} 
	}
	\IEEEauthorblockA{\textsuperscript{1}%
		Chair of Communications Engineering and Security, University of Siegen, 
		\{onur.guenlue, rafael.schaefer\}@uni-siegen.de
	}
	\IEEEauthorblockA{\textsuperscript{2}%
		Information Coding Division, Department of Electrical Engineering, Link{\"o}ping University 
	}
	\IEEEauthorblockA{\textsuperscript{3}%
		Chair of Theoretical Information Technology, Technical University of Munich, boche@tum.de}
	\IEEEauthorblockA{\textsuperscript{4}%
		CASA: Cyber Security in the Age of Large-Scale Adversaries Exzellenzcluster, Ruhr-Universit{\"a}t Bochum}
	\IEEEauthorblockA{\textsuperscript{5}%
		BMBF Research Hub 6G-Life, Technical University of Munich}
	\IEEEauthorblockA{\textsuperscript{6}%
		Department of Electrical and Computer Engineering, Princeton University, poor@princeton.edu}
}

	\maketitle

	\begin{abstract}
		The problem of secure source coding with multiple terminals is extended by considering a remote source whose noisy measurements are the correlated random variables used for secure source reconstruction. The main additions to the problem include
		\begin{inparaenum}
				\item all terminals noncausally observe a noisy measurement of the remote source;
				\item a private key is available to all legitimate terminals;
				\item the public communication link between the encoder and decoder is rate-limited; and
				\item the secrecy leakage to the eavesdropper is measured with respect to the encoder input, whereas the privacy leakage is  measured with respect to the remote source.
		\end{inparaenum}
		Exact rate regions are characterized for a lossy source coding problem with a private key, remote source, and decoder side information under security, privacy, communication, and distortion constraints. By replacing the distortion constraint with a reliability constraint, we obtain the exact rate region also for the lossless case. Furthermore, the lossy rate region for scalar discrete-time Gaussian sources and measurement channels is established.
	\end{abstract}

\IEEEpeerreviewmaketitle
\section{Introduction} \label{sec:intro}
Consider multiple terminals that observe correlated random sequences and wish to reconstruct these sequences at another terminal, called a decoder, by sending messages through noiseless communication links, i.e., the distributed source coding problem \cite{SW}. A sensor network, where each node observes a correlated random sequence that should be reconstructed at a distant node is a classic example for this problem \cite[pp.~258]{Elgamalbook}. Similarly, function computation problems in which a fusion center observes messages sent by other nodes to compute a function are closely related problems and can be used to model various recent applications \cite{CodingforComputing, benimEntropyFunctionComputation}. Since the messages sent over the communication links can be public, security constraints are imposed on these messages against an eavesdropper in the same network \cite{PrabhakaranSecureSC}. If all sent messages are available to the eavesdropper, then it is necessary to provide an advantage to the decoder over the eavesdropper to enable secure source coding. Providing side information, which is correlated with the sequences that should be reconstructed, to the decoder can provide such an advantage over the eavesdropper that can also have side information, as in \cite{GunduzErkipSecureSC,RaviSecureSC,GunduzSecureScISITDistributed}. Allowing the eavesdropper to access only a strict subset of all messages is also a method to enable secure distributed source coding, considered in \cite{LuhandKundurSecureSC,KittipongSSCwithPHelper,SalimiSecureSC}; see also \cite{MikaelSecureCEO} in which a similar method is applied to enable secure remote source reconstruction. Similarly, also a private key that is shared by legitimate terminals and hidden from the eavesdropper can provide such an advantage, as in \cite{YamamotoSecureSC,TobiasandMikaelSecureSC}.

Source coding models in the literature commonly assume that dependent multi-letter random variables are available and should be compressed. For secret-key agreement \cite{Maurer,AhlswedeCsiz} and secure function computation problems \cite{YaoSecureFunctionComp,YaoSecureFunctionComp2}, which are instances of the source coding with side information problem \cite[Section~IV-B]{OurJSAITTutorial}, the correlation between these multi-letter random variables is posited in \cite{bizimMMMMTIFS,ourISITSecurePrivateFunctArxiv} to stem from an underlying ground truth that is a remote source such that its noisy measurements are these dependent random variables. Such a remote source allows to model the cause of correlation in a network, so we also posit that there is a remote source whose noisy measurements are used in the source coding problems discussed below, which is similar to the models in \cite[pp.~78]{Bergerbook} and \cite[Fig.~9]{HaimVendingMachine}. Furthermore, in the chief executive officer (CEO) problem \cite{TheCEOFirst}, there is a remote source whose noisy measurements are encoded such that a decoder can reconstruct the remote source by using the encoder outputs. Our model is different from the model in the CEO problem, since in our model the decoder aims to recover encoder observations rather than the remote source that is considered mainly to describe the cause of correlation between encoder observations. Thus, we define the \emph{secrecy leakage} as the amount of information leaked to an eavesdropper about encoder observations. Since the remote source is common for all observations in the same network, we impose a \emph{privacy leakage} constraint on the remote source because each encoder output observed by an eavesdropper leaks information about unused encoder observations, which might later cause secrecy leakage when the unused encoder observations are employed \cite{benimdissertation, IgnaTrans, LaiTrans}; see \cite{bizimBenelux, LaiMultipleUse, benimmultientityTIFS} for joint secrecy and joint privacy constraints imposed due to multiple uses of the same source.

We characterize the rate region for a lossy secure and private source coding problem with one private key, remote source, encoder, decoder, eavesdropper, and eavesdropper and decoder side information. Requiring reliable source reconstruction, we characterize the rate region also for the lossless case. A Gaussian remote source and independent additive Gaussian noise measurement channels are considered to establish their lossy rate region under squared error distortion. 

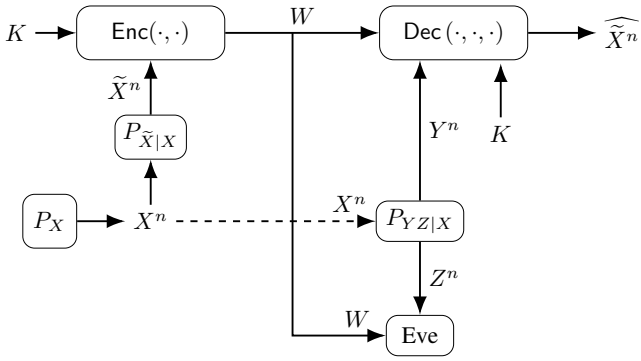
\begin{figure}
	\centering
	\resizebox{\linewidth}{!}{
		\begin{tikzpicture}
			\node (so) at (-2.5,-3.3) [draw,rounded corners = 5pt, minimum width=0.8cm,minimum height=0.8cm, align=left] {$P_X$};
			\node (a) at (-1,-0.5) [draw,rounded corners = 6pt, minimum width=2.2cm,minimum height=0.8cm, align=left] {$ \Enc(\cdot,\cdot)$};
			\node (pk) [left of = a, node distance = 2cm] {$K$};
			\node (c) at (3,-3.3) [draw,rounded corners = 5pt, minimum width=1.3cm,minimum height=0.6cm, align=left] {$P_{YZ|X}$};
			\node (f) at (-1,-2.05) [draw,rounded corners = 5pt, minimum width=1cm,minimum height=0.6cm, align=left] {$P_{\widetilde{X}|X}$};
			\node (b) at (3.5,-0.5) [draw,rounded corners = 6pt, minimum width=2.2cm,minimum height=0.8cm, align=left] {$\Dec\left(\cdot,\cdot,\cdot\right)$};
			\node (pkDec) at (4.2,-2) {$K$};
			\node (g) at (3,-5) [draw,rounded corners = 5pt, minimum width=1cm,minimum height=0.6cm, align=left] {Eve};
			\draw[decoration={markings,mark=at position 1 with {\arrow[scale=1.5]{latex}}},
			postaction={decorate}, thick, shorten >=1.4pt] (pk.east) -- (a.west);
			\draw[decoration={markings,mark=at position 1 with {\arrow[scale=1.5]{latex}}},
			postaction={decorate}, thick, shorten >=1.4pt] ($(pkDec.north)-(0.0,0)$) -- ($(pkDec.north)-(0.0,-0.8)$);
			\draw[decoration={markings,mark=at position 1 with {\arrow[scale=1.5]{latex}}},
			postaction={decorate}, thick, shorten >=1.4pt] (a.east) -- (b.west) node [midway, above] {$W$};
			\node (a1) [below of = a, node distance = 2.8cm] {$X^n$};
			\draw[decoration={markings,mark=at position 1 with {\arrow[scale=1.5]{latex}}},
			postaction={decorate}, thick, shorten >=1.4pt] ($(c.north)-(0,0)$) -- ($(b.south)-(0.5,0)$) node [midway, right] {$Y^n$};
			\draw[decoration={markings,mark=at position 1 with {\arrow[scale=1.5]{latex}}},
			postaction={decorate}, thick, shorten >=1.4pt] (so.east) -- (a1.west);
			\draw[decoration={markings,mark=at position 1 with {\arrow[scale=1.5]{latex}}},
			postaction={decorate}, thick, shorten >=1.4pt] (a1.north) -- (f.south);
			\draw[decoration={markings,mark=at position 1 with {\arrow[scale=1.5]{latex}}},
			postaction={decorate}, thick, shorten >=1.4pt] (f.north) -- (a.south) node [midway, left] {$\widetilde{X}^n$};
			\draw[decoration={markings,mark=at position 1 with {\arrow[scale=1.5]{latex}}},
			postaction={decorate}, thick, shorten >=1.4pt,dashed] (a1.east) -- ($(c.west)-(0,0.0)$) node [above  left] {$X^n$};
			\draw[decoration={markings,mark=at position 1 with {\arrow[scale=1.5]{latex}}},
			postaction={decorate}, thick, shorten >=1.4pt] (c.south) -- (g.north) node [midway, right] {$Z^n$};
			\node (b2) [right of = b, node distance = 2.5cm] {$\widehat{\widetilde{X}^n}$};
			\draw[decoration={markings,mark=at position 1 with {\arrow[scale=1.5]{latex}}},
			postaction={decorate}, thick, shorten >=1.4pt] (b.east) -- (b2.west);
			\draw[decoration={markings,mark=at position 1 with {\arrow[scale=1.5]{latex}}},
			postaction={decorate}, thick, shorten >=1.4pt] ($(a.east)+(1,0)$) -- ($(a.east)+(1,-4.35)$) -- ($(a.east)+(1,-4.50)$) -- ($(g.west)+(0,0.00)$) node [above left=0.0cm and 0.1cm of g.west] {$W$};
		\end{tikzpicture}
	}
\caption{Source coding with noisy measurements $(\widetilde{X}^n,Y^n)$ of a remote source $X^n$ and with a uniform private key $K$ under privacy, secrecy, communication, and distortion constraints.}\label{fig:ITW2022Distributed2TXSourceCodingwithSIandPrivateKey}
\end{figure}

\section{System Model}\label{sec:systemmodel}
We consider the lossy source coding model with one encoder, one decoder, and an eavesdropper (Eve), depicted in Fig.~\ref{fig:ITW2022Distributed2TXSourceCodingwithSIandPrivateKey}. The encoder $\Enc(\cdot,\cdot)$ observes a noisy measurement $\widetilde{X}^n$ of an i.i.d. remote source $X^n\sim P_X^n$ through a memoryless channel $P_{\widetilde{X}|X}$ in addition to a private key $K\in [1: 2^{nR_0}]$. The encoder output is an index $W$ that is sent over a link with limited communication rate. The decoder $\Dec(\cdot,\cdot,\cdot)$ observes the index $W$, as well as the private key $K$ and another noisy measurement $Y^n$ of the same remote source $X^n$ through another memoryless channel $P_{YZ|X}$ in order to estimate $\widetilde{X}^n$ as $\widehat{\widetilde{X}^n}$. The other noisy output $Z^n$ of $P_{YZ|X}$ is observed by Eve in addition to the index $W$. Suppose $K$ is uniformly distributed, hidden from Eve, and independent of the source output and its noisy measurements. The source and measurement alphabets are finite sets.

We next define the rate region for the lossy secure and private source coding problem defined above.

\begin{definition}\label{def:rateregionforlossySC}
	\normalfont A \emph{lossy} tuple $(R_{\text{w}},R_{\text{s}}, R_{\ell},D)\!\in\!\mathbb{R}^{4}_{\geq 0}$ is achievable, given a private key with rate $R_0\!\geq\! 0$, if for any $\delta\!>\!0$ there exist $n\!\geq\!1$, an encoder, and a decoder such that
	\begin{align}
	&\log\big|\mathcal{W}\big| \leq n(R_{\text{w}}+\delta)&&\!\!\!\!\!(\text{storage})\label{eq:storagelossySC_cons}\\
		& I(\widetilde{X}^n;W|Z^n) \leq n(R_{\text{s}}+\delta)&&\!\!\!\!\!(\text{secrecy})\label{eq:secrecyleakagelossySC_cons}\\
	&I(X^n;W|Z^n) \leq n(R_{\ell}+\delta)&&\!\!\!\!\!(\text{privacy})\label{eq:privleakagelossySC_cons}\\
& \mathbb{E}\Big[d\Big(\widetilde{X}^n,\widehat{\widetilde{X}^n}(Y^n,W,K)\Big)\Big] \leq D+\delta&&\!\!\!\!\!(\text{distortion})\label{eq:distortion_cons}
\end{align}
where $d(\widetilde{x}^n,\widehat{\widetilde{x}^n})=\frac{1}{n}\sum_{i=1}^nd(\widetilde{x}_i,\widehat{\widetilde{x}_i})$ is a per-letter bounded distortion metric. The \emph{lossy} secure and private source coding region $\mathcal{R}_{\text{D}}$ is the closure of the set of all achievable lossy tuples.\hfill $\lozenge$
\end{definition}

Note that in (\ref{eq:secrecyleakagelossySC_cons}) and (\ref{eq:privleakagelossySC_cons}) we consider conditional mutual information terms to take account of unavoidable privacy and secrecy leakages due to Eve's side information; see also \cite{LifengFCTrans,ourISITSecurePrivateFunctArxiv}. Furthermore, considering conditional mutual information terms rather than corresponding conditional entropy terms, the latter of which is used in \cite{PiantanidaSecureSC,SharagaSecureCooperativeSC,GunduzErkipSecureSC,UlukusSecureSC,TobiasandMikaelSecureSC}, to characterize the secrecy and privacy leakages simplifies our analysis.

We next define the rate region for the lossless secure and private source coding problem.

\begin{definition}\label{def:rateregionforlosslessSC}
	\normalfont A \emph{lossless} tuple $(R_{\text{w}},R_{\text{s}}, R_{\ell})\!\in\!\mathbb{R}^{3}_{\geq 0}$ is achievable, given a private key with rate $R_0\!\geq\! 0$, if for any $\delta\!>\!0$ there exist $n\!\geq\!1$, an encoder, and a decoder such that we have (\ref{eq:storagelossySC_cons})-(\ref{eq:privleakagelossySC_cons}) and 
	\begin{align}
	&&\Pr\!\Big[\widetilde{X}^n\neq \widehat{\widetilde{X}^n}(Y^n,W,K)\Big] \!\leq\! \delta&&\!\!\!\!\! (\text{reliability})\label{eq:reliablelosslessSC_cons}.
	\end{align}
	The \emph{lossless} secure and private source coding region $\mathcal{R}$ is the closure of the set of all achievable lossless tuples.\hfill $\lozenge$
\end{definition}

\section{Secure and Private Source Coding Regions}\label{sec:RateRegionsforLossyandLossless}
\subsection{Lossy Source Coding}
The lossy secure and and private source coding region $\mathcal{R}_{\text{D}}$ is characterized below; see Section~\ref{sec:ProofforLossyRegion} for its proof.

Define $[a]^-=\min\{a,0\}$ for $a\in\mathbb{R}$ and denote
\begin{align}
	R^{\prime} =[I(U;Z|V,Q)-I(U;Y|V,Q)]^-.\label{eq:definitionofRprime}
\end{align}

\begin{theorem}\label{theo:LossySC}
For given $P_X$, $P_{\widetilde{X}|X}$, $P_{YZ|X}$, and $R_0$, the region $\mathcal{R}_{\text{D}}$ is the set of all rate tuples $ (R_{\text{w}},R_{\text{s}}, R_{\ell},D)$ satisfying
	\begin{align}
	&R_{\text{w}}\geq I(U;\widetilde{X}|Y)\label{eq:RDsstorage}
	\end{align}
	and if $R_0< I(U;\widetilde{X}|Y,V)$, then
	\begin{align}
		&R_{\text{s}}\geq I(U;\widetilde{X}|Z)+R^{\prime}-R_0\label{eq:RDssecrecysmallR0}\\
		&R_{\ell}\geq I(U;X|Z)+R^{\prime}-R_0	\label{eq:RDsprivacysmallR0}
	\end{align}
	if $I(U;\widetilde{X}|Y,V)\leq R_0< I(U;\widetilde{X}|Y)$, then
	\begin{align}
		&R_{\text{s}}\geq I(V;\widetilde{X}|Z)\label{eq:RDssecrecymiddleR0}\\
		&R_{\ell}\geq I(V;X|Z)	\label{eq:RDsprivacymiddleR0}
	\end{align}
	if $R_0 \geq  I(U;\widetilde{X}|Y)$, then 
	\begin{align}
		&R_{\text{s}}\geq 0\label{eq:RDssecrecylargeR0}\\
		&R_{\ell}\geq 0	\label{eq:RDsprivacylargeR0}
	\end{align}
	for some
	\begin{align}
		P_{QVU\widetilde{X}XYZ}=P_{Q|V}P_{V|U}P_{U|\widetilde{X}}P_{\widetilde{X}|X}P_XP_{YZ|X}\label{eq:iidforlossy}
	\end{align}
	such that $\mathbb{E}\big[d\big(\widetilde{X},\widehat{\widetilde{X}}(U,Y)\big)\big] \leq D$ for some reconstruction function $\widehat{\widetilde{X}}(U,Y)$. The region $\mathcal{R}_{\text{D}}$ is convexified by using the time-sharing random variable $Q$, required  due to the $[\cdot]^-$ operation. One can limit the cardinalities to $|\mathcal{Q}|\leq 2$, $|\mathcal{V}|\leq |\widetilde{X}|+3$, and $|\mathcal{U}|\leq (|\widetilde{X}|+3)^2$.
\end{theorem}

We remark that (\ref{eq:RDssecrecylargeR0}) and (\ref{eq:RDsprivacylargeR0}) show that one can simultaneously achieve \emph{strong secrecy} and \emph{strong privacy}, i.e., the conditional mutual information terms in (\ref{eq:secrecyleakagelossySC_cons}) and (\ref{eq:privleakagelossySC_cons}), respectively, are negligible, by using a large private key $K$, which is a result missing in some recent works on secure source coding with private key.

\subsection{Lossless Source Coding}
The lossless secure and and private source coding region $\mathcal{R}$ is characterized next; see below for a proof sketch.

Denote
\begin{align}
R^{\prime\prime} =[I(\widetilde{X};Z|V,Q)-I(\widetilde{X};Y|V,Q)]^-.
\end{align}
\begin{lemma}\label{lem:LosslessSC}
	For given $P_X$, $P_{\widetilde{X}|X}$, $P_{YZ|X}$, and $R_0$, the region $\mathcal{R}$ is the set of all rate tuples $ (R_{\text{w}},R_{\text{s}}, R_{\ell})$ satisfying
	\begin{align}
	&R_{\text{w}}\geq H(\widetilde{X}|Y)\label{eq:Rsstorage}
	\end{align}
	and if $R_0< H(\widetilde{X}|Y,V)$, then
	\begin{align}
	&R_{\text{s}}\geq H(\widetilde{X}|Z)+R^{\prime\prime}-R_0\label{eq:RssecrecysmallR0}\\
	&R_{\ell}\geq I(\widetilde{X};X|Z)+R^{\prime\prime}-R_0	\label{eq:RsprivacysmallR0}
	\end{align}
	if $H(\widetilde{X}|Y,V)\leq R_0<H(\widetilde{X}|Y)$, then
	\begin{align}
	&R_{\text{s}}\geq I(V;\widetilde{X}|Z)\label{eq:RssecrecymiddleR0}\\
	&R_{\ell}\geq I(V;X|Z)	\label{eq:RsprivacymiddleR0}
	\end{align}
	if $R_0 \geq H(\widetilde{X}|Y)$, then 
	\begin{align}
	&R_{\text{s}}\geq 0\label{eq:RssecrecylargeR0}\\
	&R_{\ell}\geq 0	\label{eq:RsprivacylargeR0}
	\end{align}
	for some
	\begin{align}
	P_{QV\widetilde{X}XYZ}=P_{Q|V}P_{V|\widetilde{X}}P_{\widetilde{X}|X}P_XP_{YZ|X}\label{eq:iidforlossless}.
	\end{align}
	One can limit the cardinalities to $|\mathcal{Q}|\leq 2$ and $|\mathcal{V}|\leq |\widetilde{X}|+2$.
\end{lemma}

\begin{IEEEproof}[Proof Sketch]
	The proof for the lossless region $\mathcal{R}$ follows from the proof for the lossy region $\mathcal{R}_{\text{D}}$, given in Theorem~\ref{theo:LossySC} above, by choosing $U=\widetilde{X}$ such that we have the reconstruction function $\widehat{\widetilde{X}}(\widetilde{X},Y)=\widetilde{X}$, so we achieve $D=0$. Thus, the reliability constraint in (\ref{eq:reliablelosslessSC_cons}) is satisfied because $d(\cdot,\cdot)$ is a distortion metric.
\end{IEEEproof}

\section{Gaussian Sources and Channels}
We evaluate the lossy rate region for a Gaussian example with squared error distortion by finding the optimal auxiliary random variable in the corresponding rate region. Consider a special lossy source coding case in which 
\begin{inparaenum}[(i)]
	\item there is no private key;
	\item the eavesdropper's channel observation $Z^n$  is less noisy than the decoder's channel observation $Y^n$ 
\end{inparaenum}
such that we obtain a lossy source coding region with a single auxiliary random variable that should be optimized.

We next define less noisy channels, considering $P_{YZ|X}$.
\begin{definition}[\hspace{1sp}\cite{KoernerMartonLessNoisy}]\label{def:LN}
	\normalfont $Z$ (or eavesdropper) is \emph{less noisy} than $Y$ (or decoder) if 
	\begin{align}
	I(L;Z)\geq I(L;Y)\label{eq:condforCLN}
	\end{align}
	holds for any random variable $L$ such that $L-X-(Y,Z)$ form a Markov chain.\hfill $\lozenge$
\end{definition}

\begin{corollary}\label{cor:NoR0LessNoisy}
	For given $P_X$, $P_{\widetilde{X}|X}$, $P_{YZ|X}$, and $R_0=0$, the region $\mathcal{R}_{\text{D}}$ when the eavesdropper is less noisy than the decoder is the set of all rate tuples $ (R_{\text{w}},R_{\text{s}}, R_{\ell},D)$ satisfying
	\begin{align}
	&R_{\text{w}}\geq I(U;\widetilde{X}|Y)=I(U;\widetilde{X})-I(U;Y)\label{eq:RDsstorageforlessynoisyandnoR0}\\
	&R_{\text{s}}\geq I(U;\widetilde{X}|Z)=I(U;\widetilde{X})-I(U;Z)\label{eq:RDssecrecysmallR0forlessnoisyandnoR0}\\
	&R_{\ell}\geq I(U;X|Z)=I(U;X)-I(U;Z)	\label{eq:RDsprivacysmallR0forlessnoisyandnoR0}
	\end{align}
	for some
	\begin{align}
	P_{U\widetilde{X}XYZ}=P_{U|\widetilde{X}}P_{\widetilde{X}|X}P_XP_{YZ|X}\label{eq:iidforlossyforlessnoisyandnoR0}
	\end{align}
	such that $\mathbb{E}\big[d\big(\widetilde{X},\widehat{\widetilde{X}}(U,Y)\big)\big] \!\!\leq\!\! D$ for some reconstruction function $\widehat{\widetilde{X}}(U,Y)$. One can limit the cardinality to $|\mathcal{U}|\!\!\leq\! |\widetilde{X}|\!+\!3$.
\end{corollary}

\begin{IEEEproof}[Proof Sketch]
	The proof for Corollary~\ref{cor:NoR0LessNoisy} follows from the proof for Theorem~\ref{theo:LossySC} by considering the bounds in (\ref{eq:RDsstorage})-(\ref{eq:RDsprivacysmallR0}) since $R_0=0$. Furthermore, $R^{\prime}$ defined in (\ref{eq:definitionofRprime}) is $0$ for the less noisy condition considered, which follows because $(Q,V)-U-X-(Y,Z)$ form a Markov chain. 
\end{IEEEproof}

Suppose the following scalar discrete-time Gaussian source and channel model for the lossy source coding problem depicted in Fig.~\ref{fig:ITW2022Distributed2TXSourceCodingwithSIandPrivateKey}
\begin{align}
&X= \rho_{x}\widetilde{X}+N_{x}\label{eq:GaussianX}\\
&Y= \rho_{y}X+N_y\label{eq:GaussianY}\\
&Z=\rho_{z}X+N_z\label{eq:GaussianZ}
\end{align}
where we have the remote source $X\sim\mathcal{N}(0,1)$, fixed correlation coefficients $\rho_{x},\rho_{y},\rho_{z}\in (-1,1)$, and additive Gaussian noise random variables $N_{x}\!\sim\!\mathcal{N}(0,1\!-\!\rho_{x}^2)$, $N_{y}\!\sim\!\mathcal{N}(0,1\!-\!\rho_{y}^2)$, $N_{z}\!\sim\!\mathcal{N}(0,1\!-\!\rho_{z}^2)$ such that $(\widetilde{X},N_{x},N_y,N_z)$ are mutually independent, and we consider the squared error distortion, i.e., $d(\widetilde{x},\widehat{\widetilde{x}})\!=\!{(\widetilde{x}\!-\!\widehat{\widetilde{x}})}^2$. We remark that (\ref{eq:GaussianX}) is an inverse measurement channel $P_{X|\widetilde{X}}$ that is a weighted sum of two independent Gaussian random variables, imposed to be able to apply the conditional entropy power inequality (EPI) \cite[Lemma~II]{ConditionalEPI}; see \cite[Theorem~3]{bizimMMMMTIFS} and \cite[Section~V]{CorrelatedPaperLong} for binary symmetric inverse channel assumptions imposed to apply Mrs. Gerber's lemma \cite{WZGerber}. Suppose $|\rho_{z}|>|\rho_{y}|$ such that $Y$ is stochastically-degraded than $Z$ since then there exists a random variable $\widetilde{Y}$ such that $P_{\widetilde{Y}|X}=P_{Y|X}$ and $P_{\widetilde{Y}Z|X}\!=\!P_{Z|X}P_{\widetilde{Y}|Z}$ \cite[Lemma~6]{ShunGaussiandegradation}, so $Z$ is also less noisy than $Y$ since less noisy channels constitute a strict superset of the set of stochastically-degraded channels and both channel sets consider only the conditional marginal probability distributions \cite[pp.~121]{Elgamalbook}.

We next take the liberty to use the lossy rate region in Corollary~\ref{cor:NoR0LessNoisy}, characterized for discrete memoryless channels, for the model in (\ref{eq:GaussianX})-(\ref{eq:GaussianZ}). This is common in the literature since there is a discretization procedure to extend the achievability proof to well-behaved continuous-alphabet random variables and the converse proof applies to arbitrary random variables; see \cite[Remark~3.8]{Elgamalbook}. For Gaussian sources and channels, we use differential entropy and eliminate the cardinality bound on the auxiliary random variable. The lossy source coding region for the model in (\ref{eq:GaussianX})-(\ref{eq:GaussianZ}) without a private key is given below.

\begin{lemma}\label{lem:Gaussianregion}
	For the model in (\ref{eq:GaussianX})-(\ref{eq:GaussianZ}) such that $|\rho_{z}|>|\rho_{y}|$ and $R_0=0$, the region $\mathcal{R}_{\text{D}}$ with squared error distortion is the set of all rate tuples $ (R_{\text{w}},R_{\text{s}}, R_{\ell},D)$ satisfying, for $0<\alpha\leq 1$,
	\begin{align}
	&R_{\text{w}}\geq \frac{1}{2}\log\Big(\frac{1-\rho_x^2\rho_y^2(1-\alpha)}{\alpha}\Big)\label{eq:RDsstorageforlessynoisyandnoR0Gauss}\\
	&R_{\text{s}}\geq \frac{1}{2}\log \Big( \frac{1-\rho_x^2\rho_z^2(1-\alpha)}{\alpha}\Big) \label{eq:RDssecrecysmallR0forlessnoisyandnoR0Gauss}\\
	&R_{\ell}\geq 	\frac{1}{2}\log \Big(\frac{1-\rho_x^2\rho_z^2(1-\alpha)}{1-\rho_x^2(1-\alpha)}\Big)\label{eq:RDsprivacysmallR0forlessnoisyandnoR0Gauss}\\
	&D\geq \frac{\alpha(1-\rho_{x}^2\rho_y^2)}{1-\rho_x^2\rho_y^2(1-\alpha)}.\label{eq:RDsdistortionsmallR0forlessnoisyandnoR0Gauss}
	\end{align}
\end{lemma}

\begin{IEEEproof}[Proof Sketch]
	For the achievability proof, let $U\sim \mathcal{N}(0,1\!-\!\alpha)$ and $\Theta\sim \mathcal{N}(0,\alpha)$, as in \cite[Eq.~(34)]{FransGaussian} and \cite[Appendix~B]{HidekiYagiGaussianArxiv}, be independent random variables for some $0<\alpha\leq 1$ such that $\widetilde{X}=U+\Theta$ and $U-\widetilde{X}-X-(Y,Z)$ form a Markov chain. Choose the reconstruction function $\widehat{\widetilde{X}}(U,Y)$ as the minimum mean square error (MMSE) estimator, and given any fixed $D>0$ auxiliary random variables are chosen such that the distortion constraint is satisfied. We then have for the squared error distortion
	\begin{align}
		D= \mathbb{E}\Big[\big(\widetilde{X}-\widehat{\widetilde{X}}(U,Y)\big)^2\Big] \overset{(a)}{=}\frac{1}{2\pi e}e^{2h(\widetilde{X}|U,Y)}\label{eq:definitionofDasMMSE}
	\end{align}
	where equality in $(a)$ is achieved because $\widetilde{X}$ is Gaussian and the reconstruction function is the MMSE estimator \cite[Theorem~8.6.6]{CoverandThomas}. Define the covariance matrix of the vector random variable $[\widetilde{X}, U, Y]$ as $\mathbf{K}_{\widetilde{X}UY}$ and of $[U, Y]$ as $\mathbf{K}_{UY}$, respectively. We then have 
	\begin{align}
			&h(\widetilde{X}|U,Y)=h(\widetilde{X},U,Y)-h(U,Y) \nonumber\\
			&= \frac{1}{2}\log \bigg(2\pi e \frac{\text{det}(\mathbf{K}_{\widetilde{X}UY})}{\text{det}(\mathbf{K}_{UY})}\bigg)\label{eq:hXtildegivenUY}
	\end{align}
	where $\text{det}(\cdot)$ is the determinant of a matrix; see also \cite[Section~F]{MikaelSecureCEO}. Combining (\ref{eq:definitionofDasMMSE}) and (\ref{eq:hXtildegivenUY}), and calculating the determinants, we obtain 
	\begin{align}
		D= \frac{\alpha(1-\rho_{x}^2\rho_y^2)}{1-\rho_x^2\rho_y^2(1-\alpha)}.\label{eq:achdefD}
	\end{align}
	One can also show that 
	\begin{align}
		&I(U;\widetilde{X})\!=\!h(\widetilde{X})\!-\!h(\widetilde{X}|U)\!=\!\frac{1}{2}\log\Big(\frac{1}{\alpha}\Big) \\
		&I(U;X)\! =\!h(X)\!-\!h(X|U)\!=\! \frac{1}{2}\log\Big(\frac{1}{1-\rho_x^2(1-\alpha)}\Big)\\
		& I(U;Y)\!=\! h(Y)\!-\!h(Y|U)\!=\!\frac{1}{2}\log \Big(\frac{1}{1-\rho_x^2\rho_y^2(1-\alpha)}\Big)\\
		&I(U;Z)\!=\!  h(Z)\!-\!h(Z|U)\!=\!\frac{1}{2}\log \Big(\frac{1}{1-\rho_x^2\rho_z^2(1-\alpha)}\Big).
	\end{align}
	Thus, by calculating (\ref{eq:RDsstorageforlessynoisyandnoR0})-(\ref{eq:RDsprivacysmallR0forlessnoisyandnoR0}), the achievability proof follows.
	
	For the converse proof, one can first show that
	\begin{align}
		&I(U;\widetilde{X})-I(U;Y)= h(Y|U)-h(\widetilde{X}|U)\label{eq:converseIUXtildeYopenfirst}\\
		&I(U;\widetilde{X})-I(U;Z)= h(Z|U)-h(\widetilde{X}|U)\\
		&I(U;X)-I(U;Z)= h(Z|U)-h(X|U)
	\end{align}
	which follow since $h(\widetilde{X})=h(X)=h(Y)=h(Z)$. Suppose
	\begin{align}
		&h(\widetilde{X}|U) =\frac{1}{2}\log (2\pi e\alpha)\label{eq:conversehxtildeuy}
	\end{align}
		for any $0<\alpha\leq 1$ that represents the unique variance of a Gaussian random variable; see \cite[Lemma~2]{bizimMMMMTIFS} for a similar result applied to binary random variables. Thus, by applying the conditional EPI, we obtain 
	\begin{align}
	&e^{2h(Y|U)} \overset{(a)}{=}e^{2h(\rho_x\rho_y\widetilde{X}|U)} +e^{2h(\rho_yN_x+N_y)}\nonumber\\
	&=2\pi e \big(\rho_x^2\rho_y^2\alpha+ \rho_y^2(1-\rho_x^2)+1-\rho_y^2\big)\nonumber\\
	&=2\pi e \big(1-\rho_x^2\rho_y^2(1-\alpha)\big)\label{eq:convhYgivenU}
	\end{align}
	where $(a)$ follows because $U-\widetilde{X}-(N_x,N_y)$ form a Markov chain and $(N_x,N_y)$ are independent of $\widetilde{X}$, so $(N_x,N_y)$ are independent of $U$, and equality is satisfied since, given $U$, $\rho_x\rho_y\widetilde{X}$ and $(\rho_yN_x+N_y)$ are conditionally independent and they are Gaussian random variables, as imposed in (\ref{eq:conversehxtildeuy}) above; see \cite[Lemma~1 and Eq. (28)]{bizimMMMMTIFS} for a similar result applied to binary random variables by extending Mrs. Gerber's lemma. Similarly, we have
	\begin{align}
		&e^{2h(Z|U)} =2\pi e \big(1-\rho_x^2\rho_z^2(1-\alpha)\big)
	\end{align}
	which follows by replacing $(Y,\rho_y,N_y)$ with $(Z, \rho_z,N_z)$ in (\ref{eq:convhYgivenU}), respectively, because the channel $P_{Y|U}$ can be mapped to $P_{Z|U}$ with these changes due to (\ref{eq:GaussianX})-(\ref{eq:GaussianZ}) and the Markov chain $U-\widetilde{X}-X-(Y,Z)$. Furthermore, we have
	\begin{align}
			&e^{2h(X|U)} \overset{(a)}{=}e^{2h(\rho_x\widetilde{X}|U)} +e^{2h(N_x)}\nonumber\\
			&=2\pi e \big(\rho_x^2\alpha+1-\rho_x^2\big)\nonumber\\
			&=2\pi e \big(1-\rho_x^2(1-\alpha)\big)\label{eq:convhXgivenU}
	\end{align}
	where $(a)$ follows because $N_x$ is independent of $U$, and equality is achieved since, given $U$, $\rho_x\widetilde{X}$ and $N_x$ are conditionally independent and are Gaussian random variables. Therefore, by applying (\ref{eq:converseIUXtildeYopenfirst})-(\ref{eq:convhXgivenU}) to (\ref{eq:RDsstorageforlessynoisyandnoR0})-(\ref{eq:RDsprivacysmallR0forlessnoisyandnoR0}), the converse proof for (\ref{eq:RDsstorageforlessynoisyandnoR0Gauss})-(\ref{eq:RDsprivacysmallR0forlessnoisyandnoR0Gauss}) follows.
	
	Next, consider
	\begin{align}
		&h(\widetilde{X}|U,Y)=-I(U;\widetilde{X}|Y)+h(\widetilde{X}|Y)\nonumber\\
		&\overset{(a)}{=}-h(Y|U)+h(\widetilde{X}|U)+h(Y|\widetilde{X})\nonumber\\
		&\overset{(b)}{=}\frac{1}{2}\log\Big(\frac{\alpha}{1-\rho_x^2\rho_y^2(1-\alpha)}\Big)+ h(\rho_x\rho_y\widetilde{X}\!+\!\rho_yN_x\!+\!N_y|\widetilde{X})\nonumber\\
		&\overset{(c)}{=}\frac{1}{2}\log\Big(\frac{\alpha}{1-\rho_x^2\rho_y^2(1-\alpha)}\Big)+ h(\rho_yN_x\!+\!N_y)\nonumber\\
		&=\frac{1}{2}\log\Big(2\pi e\frac{\alpha (\rho_y^2(1-\rho_x^2)+(1-\rho_y^2))}{1-\rho_x^2\rho_y^2(1-\alpha)}\Big)\nonumber\\
		&=\frac{1}{2}\log\Big(2\pi e\frac{\alpha (1-\rho_x^2\rho_y^2)}{1-\rho_x^2\rho_y^2(1-\alpha)}\Big)\label{eq:conversehXtildeUYequality}
	\end{align}
	where $(a)$ follows by (\ref{eq:RDsstorageforlessynoisyandnoR0}) and (\ref{eq:converseIUXtildeYopenfirst}), and since $h(Y)=h(\widetilde{X})$, $(b)$ follows by (\ref{eq:conversehxtildeuy}) and (\ref{eq:convhYgivenU}), and $(c)$ follows because $(N_x,N_y)$ are independent of $\widetilde{X}$. Furthermore, for any random variable $\widetilde{X}$ and reconstruction function $\widehat{\widetilde{X}}(U,Y)$, we have \cite[Theorem~8.6.6]{CoverandThomas}
	\begin{align}
		\mathbb{E}\Big[\big(\widetilde{X}-\widehat{\widetilde{X}}(U,Y)\big)^2\Big] \geq \frac{1}{2\pi e}e^{2h(\widetilde{X}|U,Y)}.\label{eq:conversedistortionlowerbyhXtildeUY}
	\end{align}
	Combining the distortion constraint given in Corollary~\ref{cor:NoR0LessNoisy} with (\ref{eq:conversehXtildeUYequality}) and (\ref{eq:conversedistortionlowerbyhXtildeUY}), the converse proof for (\ref{eq:RDsdistortionsmallR0forlessnoisyandnoR0Gauss}) follows.
\end{IEEEproof}

\section{Proof for Theorem~\ref{theo:LossySC}}\label{sec:ProofforLossyRegion}
\subsection{Achievability Proof for Theorem~\ref{theo:LossySC}}
\begin{IEEEproof}[Proof Sketch]
	We leverage the output statistics of random binning (OSRB) method \cite{AhlswedeCsiz,OSRBAmin,RenesRenner} for the achievability proof by following the steps described in \cite[Section~1.6]{BlochLectureNotes2018}. 
	
	Let $(V^n,U^n,\widetilde{X}^n,X^n,Y^n,Z^n)$ be i.i.d. according to $P_{VU\widetilde{X}XYZ}$ that can be obtained from (\ref{eq:iidforlossy}) by fixing $P_{U|\widetilde{X}}$ and $P_{V|U}$ such that $\mathbb{E}[d\big(\widetilde{X},\widehat{\widetilde{X}})]\leq (D+\epsilon)$ for any $\epsilon>0$. To each $v^n$ assign two random bin indices $F_{\text{v}}\in[1:2^{n\widetilde{R}_{\text{v}}}]$ and $W_{\text{v}}\in[1:2^{nR_{\text{v}}}]$. Furthermore, to each $u^n$ assign three random bin indices $F_{\text{u}}\in[1:2^{n\widetilde{R}_{\text{u}}}]$, $W_{\text{u}}\in[1:2^{nR_{\text{u}}}]$, and $K_{\text{u}}\in[1:~2^{nR_0}]$, where $R_0$ is the private key rate defined in Section~\ref{sec:systemmodel}. The public indices $F=(F_{\text{v}},F_{\text{u}})$ represent the choice of a source encoder and decoder pair. Furthermore, we impose that the messages sent by the source encoder $\Enc(\cdot,\cdot)$ to the source decoder $\Dec(\cdot,\cdot,\cdot)$ are
	\begin{align}
	&W=(W_{\text{v}}, W_{\text{u}}, K+K_{\text{u}})\label{eq:LossyAchW}
	\end{align}
	where the summation with the private key is in modulo-~$2^{nR_0}$, i.e., one-time padding.

	The public index $F_{\text{v}}$ is almost independent of $(\widetilde{X}^n,X^n,Y^n,Z^n)$ if we have \cite[Theorem 1]{OSRBAmin}
	\begin{align}
		\widetilde{R}_{\text{v}}<H(V|\widetilde{X},X,Y,Z)\overset{(a)}{=}H(V|\widetilde{X})\label{eq:independenceofFv}
	\end{align} 
	where $(a)$ follows since $(X,Y,Z)-\widetilde{X}-V$ form a Markov chain. The constraint in (\ref{eq:independenceofFv}) suggests that the expected value, taken over the random bin assignments, of the variational distance between the joint probability distributions $\text{Unif}[1\!\!:\!2^{n\widetilde{R}_{\text{v}}}]\cdot~ P_{\widetilde{X}^n}$ and $P_{F_{\text{v}}\widetilde{X}^n}$ vanishes when $n\rightarrow\infty$. Moreover, the public index $F_{\text{u}}$ is almost independent of $(V^n,\widetilde{X}^n,X^n,Y^n,Z^n)$ if we have
	\begin{align}
		\widetilde{R}_{\text{u}}<H(U|V,\widetilde{X},X,Y,Z)\overset{(a)}{=}H(U|V,\widetilde{X})\label{eq:independenceofFu}
	\end{align}
	where $(a)$ follows from the Markov chain $(X,Y,Z)-\widetilde{X}-(U,V)$. 
	
	Using a Slepian-Wolf (SW) \cite{SW} decoder that observes $(Y^n,F_{\text{v}},W_{\text{v}})$, one can reliably estimate $V^n$ if we have \cite[Lemma 1]{OSRBAmin}
	\begin{align}
		\widetilde{R}_{\text{v}} + R_{\text{v}}> H(V|Y)\label{eq:Vnreconstruction}
	\end{align}	
	since then the expected error probability, taken over random bin assignments, vanishes when $n\rightarrow\infty$. Furthermore, one can reliably estimate $U^n$ by using a SW decoder that observes $(K, V^n,Y^n,F_{\text{u}},W_{\text{u}},K+K_{\text{u}})$ if we have
	\begin{align}
		&R_0+\widetilde{R}_{\text{u}} + R_{\text{u}}> H(U|V,Y)\label{eq:Unreconstruction}.
	\end{align}

	To satisfy (\ref{eq:independenceofFv})-(\ref{eq:Unreconstruction}), for any $\epsilon>0$ we fix 
	\begin{align}
	&\widetilde{R}_{\text{v}} = H(V|\widetilde{X})-\epsilon\label{eq:R_vtildechosen}\\
	&R_{\text{v}} = I(V;\widetilde{X})-I(V;Y)+2\epsilon\label{eq:R_vchosen}\\
	&\widetilde{R}_{\text{u}} = H(U|V,\widetilde{X})-\epsilon\label{eq:R_utildechosen}\\
	&R_0+R_{\text{u}} = I(U;\widetilde{X}|V)-I(U;Y|V)+2\epsilon\label{eq:R_uchosen}.
	\end{align}
	Since all tuples $(v^n,u^n,\widetilde{x}^n,x^n,y^n,z^n)$ are in the jointly typical set with high probability, by the typical average lemma \cite[pp. 26]{Elgamalbook}, the distortion constraint (\ref{eq:distortion_cons}) is satisfied.

	\textbf{Communication Rate}: (\ref{eq:R_vchosen}) and (\ref{eq:R_uchosen}) result in a communication (storage) rate of
	\begin{align}
	&R_{\text{w}} = R_0+R_{\text{v}} + R_{\text{u}} \overset{(a)}{=}  I(U;\widetilde{X}|Y)+4\epsilon \label{eq:R_wchosen}
	\end{align}
	where $(a)$ follows since $V-U-\widetilde{X}-Y$ form a Markov chain. 

	\textbf{Privacy Leakage Rate}: Since the private key $K$ is uniformly distributed and is independent of source and channel random variables, we can consider the following virtual scenario to calculate the leakage. We first assume for the virtual scenario that there is no private key such that the encoder output for the virtual scenario is
	\begin{align}
	&\widebar{W}=(W_{\text{v}}, W_{\text{u}}, K_{\text{u}})\label{eq:AchWbar}.
	\end{align}
	We calculate the leakage for the virtual scenario. Then, given the mentioned properties of the private key and due to the one-time padding step in (\ref{eq:LossyAchW}), we can subtract $H(K)=nR_0$ from the leakage calculated for the virtual scenario to obtain the leakage for the original problem, which follows from the sum of (\ref{eq:R_utildechosen}) and (\ref{eq:R_uchosen}) if $\epsilon\rightarrow 0$ when $n\rightarrow\infty$. Thus, we have the privacy leakage 
	\begin{align}
	&I(X^n;W,F|Z^n)=I(X^n;\widebar{W},F|Z^n)-nR_0\nonumber\\
	&\overset{(a)}{=}\!H(\widebar{W},F|Z^n)\!-\!H(\widebar{W},F|X^n)\!-\!nR_0\nonumber\\
	&\overset{(b)}{=} H(\widebar{W},F|Z^n) -H(U^n,V^n|X^n)\nonumber\\
	&\qquad + H(V^n|\widebar{W},F,X^n) + H(U^n|V^n,\widebar{W},F,X^n)-nR_0\nonumber\\
	&\overset{(c)}{\leq} H(\widebar{W},F|Z^n) -nH(U,V|X) +2n\epsilon_n-nR_0\label{eq:ach_privtoEvefirststep}
	\end{align}
	where $(a)$ follows because $(\widebar{W},F)-X^n-Z^n$ form a Markov chain, $(b)$ follows since $(U^n,V^n)$ determine $(F_{\text{u}},W_{\text{u}},K_{\text{u}},F_{\text{v}},W_{\text{v}})$, and $(c)$ follows since $(U^n,V^n,X^n)$ is i.i.d. and for some $\epsilon_n>0$ such that $\epsilon_n\rightarrow0$ when $n\rightarrow\infty$ because $(F_{\text{v}},W_{\text{v}},X^n)$ can reliably recover $V^n$ by (\ref{eq:Vnreconstruction}) because of the Markov chain $V^n-X^n-Y^n$ and, similarly, $(F_{\text{u}},W_{\text{u}},K_{\text{u}},V^n,X^n)$ can reliably recover $U^n$ by (\ref{eq:Unreconstruction}) because of  $H(U|V,Y)\geq H(U|V,X)$ that is proved in \cite[Eq. (55)]{ourISITSecurePrivateFunctArxiv} for the Markov chain $(V,U)-X-Y$. 

	Next, we consider the term $H(\widebar{W},F|Z^n)$ in (\ref{eq:ach_privtoEvefirststep}) and provide single letter bounds on it by applying the six different decodability results given in \cite[Section~V-A]{ourISITSecurePrivateFunctArxiv} that are applied to an entirely similar conditional entropy term in \cite[Eq. (54)]{ourISITSecurePrivateFunctArxiv} that measures the uncertainty in indices conditioned on an i.i.d. multi-letter random variable. Thus, combining the six decodability results in \cite[Section~V-A]{ourISITSecurePrivateFunctArxiv} with (\ref{eq:ach_privtoEvefirststep}) we obtain
   \begin{align}
		&I(X^n;W,F|Z^n)\leq n\big([I(U;Z|V)-I(U;Y|V)+\epsilon]^-\nonumber\\
		&\qquad\qquad\qquad\qquad\qquad+I(U;X|Z)+3\epsilon_n-R_0\big)\label{eq:privEveachfin}.
    \end{align}

	We remark that (\ref{eq:R_uchosen}) implicitly assumes that the private key rate $R_0$ is less than $(I(U;\widetilde{X}|V)-I(U;Y|V)+~2\epsilon)=(I(U;\widetilde{X}|Y,V)+2\epsilon)$, where the equality follows from the Markov chain $(V,U)-\widetilde{X}-Y$. The communication rate results are not affected by this assumption since $\widetilde{X}^n$ should be reconstructed by the decoder. However, if the private key rate $R_0$ is greater than or equal to $(I(U;\widetilde{X}|Y,V)+2\epsilon)$, then we can remove the bin index $K_{\text{u}}$ from the code construction above and apply one-time padding to the bin index $W_{\text{u}}$ such that we have the encoder output 
	\begin{align}
	&\widebar{\widebar{W}}=(W_{\text{v}}, W_{\text{u}}+K)\label{eq:AchWbarbar}
	\end{align}
	where the summation with the private key is in modulo-~$2^{nR_{\text{u}}}=2^{n(I(U;\widetilde{X}|Y,V)+2\epsilon)}$.	Thus, one then does not leak any information about $W_{\text{u}}$ to the eavesdropper because of the one-time padding step in (\ref{eq:AchWbarbar}). We then have the privacy leakage
	\begin{align}
		&I(X^n;\widebar{\widebar{W}},F|Z^n)=I(X^n;W_{\text{v}},F|Z^n)\nonumber\\
		&\overset{(a)}{\leq} H(X^n|Z^n)-H(X^n|Z^n,W_{\text{v}},F_{\text{v}})+\epsilon^{\prime}_n\nonumber\\
		&\overset{(b)}{\leq} H(X^n|Z^n)-H(X^n|Z^n,V^n)+\epsilon^{\prime}_n\nonumber\\
		&\overset{(c)}{=}nI(V;X|Z)+\epsilon^{\prime}_n\label{eq:AchprivWbarbar}
	\end{align}
	where $(a)$ follows for some $\epsilon^{\prime}_n$ such that $\epsilon^{\prime}_n\rightarrow0$ when $n\rightarrow\infty$ since by (\ref{eq:independenceofFu}) $F_{\text{u}}$ is almost independent of $(V^n,X^n,Z^n)$; see also \cite[Theorem~1]{HolensteinRenner}, $(b)$ follows since $V^n$ determines $(F_{\text{v}},W_{\text{v}})$, and $(c)$ follows because $(X^n,Z^n,V^n)$ are i.i.d. 
	
	Note that we can reduce the privacy leakage given in (\ref{eq:AchprivWbarbar}) if $R_0\geq (I(U;\widetilde{X})-I(U;Y)+4\epsilon)=(I(U;\widetilde{X}|Y)+4\epsilon)$, where the equality follows from the Markov chain $U-\widetilde{X}-Y$, since then we can apply one-time padding to both bin indices $W_{\text{v}}$ and $W_{\text{u}}$ with the sum rate
	\begin{align}
		&R_{\text{v}}+R_{\text{u}}\nonumber\\
		& \overset{(a)}{=} I(V;\widetilde{X})-I(V;Y)+2\epsilon \!+\! I(U;\widetilde{X}|V)\!-\!I(U;Y|V)\!+\!2\epsilon\nonumber\\
		&\overset{(b)}{=}I(U;\widetilde{X})-I(U;Y)+4\epsilon 
	\end{align}
	where $(a)$ follows by (\ref{eq:R_vchosen}) and (\ref{eq:R_uchosen}), and $(b)$ follows from the Markov chain $V-U-\widetilde{X}-Y$. Thus, one then does not leak any information about $(W_{\text{v}},W_{\text{u}})$ to the eavesdropper because of the one-time padding step, so we then obtain the privacy leakage of
	\begin{align}
		&I(X^n;F|Z^n)=I(X^n;F_{\text{v}}|Z^n)+I(X^n;F_{\text{u}}|Z^n,F_{\text{v}}) \nonumber\\
		&\overset{(a)}{\leq} 2\epsilon^{\prime}_n\label{eq:Achprivacyleakzero}
	\end{align}
	where $(a)$ follows since by (\ref{eq:independenceofFv}) $F_{\text{v}}$ is almost independent of $(X^n,Z^n)$ and	by (\ref{eq:independenceofFu}) $F_{\text{u}}$ is almost independent of $(V^n,X^n,Z^n)$.

\textbf{Secrecy Leakage Rate}: Similar to the privacy leakage analysis above, we first consider the virtual scenario with the encoder output given in (\ref{eq:AchWbar}), and then calculate the leakage for the original problem by subtracting $H(K)=nR_0$ from the leakage calculated for the virtual scenario. Thus, we obtain
\begin{align}
&I(\widetilde{X}^n;W,F|Z^n)=I(\widetilde{X}^n;\widebar{W},F|Z^n)-nR_0\nonumber\\
&\overset{(a)}{=}H(\widebar{W},F|Z^n)-H(\widebar{W},F|\widetilde{X}^n)-nR_0\nonumber\\
&\overset{(b)}{=} H(\widebar{W},F|Z^n) -H(U^n,V^n|\widetilde{X}^n)\nonumber\\
&\qquad + H(V^n|\widebar{W},F,\widetilde{X}^n)+ H(U^n|V^n,\widebar{W},F,\widetilde{X}^n)\nonumber\\
&\overset{(c)}{\leq} H(\widebar{W},F|Z^n) -nH(U,V|\widetilde{X})+2n\epsilon^{\prime}_n-nR_0\nonumber\\
&\overset{(d)}{\leq}n\big([I(U;Z|V)-I(U;Y|V)+\epsilon]^-\nonumber\\
&\qquad\qquad+I(U;\widetilde{X}|Z)+3\epsilon_n^{\prime}-R_0\big)\label{eq:ach_secrecyfirststep}
\end{align}
where $(a)$ follows from the Markov chain $(\widebar{W},F)-\widetilde{X}^n-Z^n$, $(b)$ follows since $(U^n,V^n)$ determine $(\widebar{W},F)$, $(c)$ follows because $(V^n,U^n,\widetilde{X}^n)$ are i.i.d. and because $(F_{\text{v}},W_{\text{v}},\widetilde{X}^n)$ can reliably recover $V^n$ by (\ref{eq:Vnreconstruction}) due to the Markov chain $V^n-\widetilde{X}^n-Y^n$ and, similarly, $(F_{\text{u}},W_{\text{u}},K_{\text{u}},V^n,\widetilde{X}^n)$ can reliably recover $U^n$ by (\ref{eq:Unreconstruction}) due to  $H(U|V,Y)\!\geq\! H(U|V,\widetilde{X})$ that can be proved as in \cite[Eq. (55)]{ourISITSecurePrivateFunctArxiv} for the Markov chain $(V,U)-\widetilde{X}-Y$, and $(d)$ follows by applying the six decodability results in \cite[Section~V-A]{ourISITSecurePrivateFunctArxiv} that are applied to (\ref{eq:ach_privtoEvefirststep}) with the final result in (\ref{eq:privEveachfin}) by replacing $X$ with $\widetilde{X}$.

Similar to the privacy leakage analysis above if we have $R_0\geq (I(U;\widetilde{X}|Y,V)+2\epsilon)$, then we can eliminate $K_{\text{u}}$ and apply one-time padding as in (\ref{eq:AchWbarbar}) such that no information about $W_{\text{u}}$ is leaked to the eavesdropper and we have
	\begin{align}
&I(\widetilde{X}^n;\widebar{\widebar{W}},F|Z^n)=I(\widetilde{X}^n;W_{\text{v}},F|Z^n)\nonumber\\
&\overset{(a)}{\leq} H(\widetilde{X}^n|Z^n)-H(\widetilde{X}^n|Z^n,W_{\text{v}},F_{\text{v}})+\epsilon^{\prime}_n\nonumber\\
&\overset{(b)}{\leq} H(\widetilde{X}^n|Z^n)-H(\widetilde{X}^n|Z^n,V^n)+\epsilon^{\prime}_n\nonumber\\
&\overset{(c)}{=}nI(V;\widetilde{X}|Z)+\epsilon^{\prime}_n\label{eq:AchsecrecyWbarbar}
\end{align}
where $(a)$ follows because by (\ref{eq:independenceofFu}) $F_{\text{u}}$ is almost independent of $(V^n,\widetilde{X}^n,Z^n)$, $(b)$ follows since $V^n$ determines $(F_{\text{v}},W_{\text{v}})$, and $(c)$ follows because $(\widetilde{X}^n,Z^n,V^n)$ are i.i.d.

If $R_0\geq (I(U;\widetilde{X}|Y)+4\epsilon)$, we can apply one-time padding to hide $(W_{\text{v}},W_{\text{u}})$, as in the privacy leakage analysis above. We then have the secrecy leakage of 
\begin{align}
	&I(\widetilde{X}^n;F|Z^n)=I(\widetilde{X}^n;F_{\text{v}}|Z^n)+I(\widetilde{X}^n;F_{\text{u}}|Z^n,F_{\text{v}}) \nonumber\\
	&\overset{(a)}{\leq} 2\epsilon^{\prime}_n\label{eq:Achsecrecyleakzero}
\end{align}
where $(a)$ follows since by (\ref{eq:independenceofFv}) $F_{\text{v}}$ is almost independent of $(\widetilde{X}^n,Z^n)$ and by (\ref{eq:independenceofFu}) $F_{\text{u}}$ is almost independent of $(V^n,\widetilde{X}^n,Z^n)$.

Suppose the public indices $F$ are generated uniformly at random, and the encoder generates $(V^n,U^n)$ according to $P_{V^nU^n|\widetilde{X}^nF_{\text{v}}F_{\text{u}}}$ that can be obtained from the proposed binning scheme above to compute the bins $W_{\text{v}}$ from $V^n$ and $W_{\text{u}}$ from $U^n$, respectively. Such a procedure results in a joint probability distribution almost equal to $P_{VU\widetilde{X}XYZ}$ fixed above \cite[Section 1.6]{BlochLectureNotes2018}. Note that the privacy and secrecy leakage metrics above are expectations over all possible public index realizations $F=f$. Therefore, using a time-sharing random variable $Q$ for convexification and applying the selection lemma \cite[Lemma 2.2]{Blochbook} to each decodability case separately, the achievability for Theorem~\ref{theo:LossySC} follows by choosing an $\epsilon>0$ such that $\epsilon\rightarrow 0$ when $n\rightarrow\infty$.
\end{IEEEproof}
 
\subsection{Converse Proof for Theorem~\ref{theo:LossySC}}
\begin{IEEEproof}[Proof Sketch]
Assume that for some $\delta_n\!>\!0$ and $n\geq 1$, there exist an encoder and a decoder such that (\ref{eq:storagelossySC_cons})-(\ref{eq:distortion_cons}) are satisfied for some tuple $(R_{\text{w}}, R_{\text{s}},R_{\ell},D)$ given a private key with rate $R_0$. 

Define $V_{i}\triangleq (W,Y^{n}_{i+1},Z^{i-1})$ and $U_{i}\triangleq(W,Y^{n}_{i+1},Z^{i-1},X^{i-1},K)$ that satisfy the Markov chain $V_{i}-U_{i}-\widetilde{X}_{i}-X_i-(Y_{i},Z_{i})$ by definition of the source statistics. We have 
\begin{align}
&D+\delta_n \overset{(a)}{\geq} \mathbb{E}\Big[d\Big(\widetilde{X}^n,\widehat{\widetilde{X}^n}(Y^n,W,K)\Big)\Big] \nonumber\\
&\overset{(b)}{\geq} \mathbb{E}\Big[d\Big(\widetilde{X}^n,\widehat{\widetilde{X}^n}(Y^n,W,K,X^{i-1},Z^{i-1})\Big)\Big]\nonumber\\
&\overset{(c)}{=}\mathbb{E}\Big[d\Big(\widetilde{X}^n,\widehat{\widetilde{X}^n}(Y_{i}^n,W,K,X^{i-1},Z^{i-1})\Big)\Big]\nonumber\\
&\overset{(d)}{=}\frac{1}{n}\sum_{i=1}^{n} \mathbb{E}\Big[d\Big(\widetilde{X}_i,\widehat{\widetilde{X}_i}(U_i,Y_i)\Big)\Big] \label{eq:conversedistortion}
\end{align}
where $(a)$ follows by (\ref{eq:distortion_cons}), $(b)$ follows since providing more information to the reconstruction function does not increase expected distortion, $(c)$ follows from the Markov chain 
\begin{align}
Y^{i-1}-(Y_i^n,X^{i-1},Z^{i-1},W,K)-\widetilde{X}^n
\end{align}
and $(d)$ follows from the definition of $U_i$.

\textbf{Communication Rate}: For any $R_0\geq 0$, we have
\begin{align}
&n(R_{\text{w}}+\delta_n) \overset{(a)}{\geq} \log|\mathcal{W}|\nonumber\\
&\geq H(W|Y^n,K)-H(W|Y^n,K,\widetilde{X}^n)\nonumber\\
&\overset{(b)}{=}\sum_{i=1}^n I(W;\widetilde{X}_i|\widetilde{X}^{i-1},Y_{i+1}^{n},Z^{i-1},K,Y_i)\nonumber\\
&\overset{(c)}{=}\sum_{i=1}^n I(\widetilde{X}^{i-1},Y_{i+1}^{n},Z^{i-1},K,W;\widetilde{X}_i|Y_i)\nonumber\\
&\overset{(d)}{\geq}\sum_{i=1}^n I(X^{i-1},Y_{i+1}^{n},Z^{i-1},K,W;\widetilde{X}_i|Y_i)\nonumber\\
&=\sum_{i=1}^n I(U_i;\widetilde{X}_i|Y_i)
\end{align}
where $(a)$ follows by (\ref{eq:storagelossySC_cons}), $(b)$ follows  from the Markov chain
\begin{align}
(Y^{i-1},X^{i-1},Z^{i-1})-(\widetilde{X}^{i-1},Y_i^n,K)-(\widetilde{X}_i,W)\label{eq:intermMarkov}
\end{align}
$(c)$ follows because $(\widetilde{X}_i,Y_i)$ are independent of $(\widetilde{X}^{i-1},Y_{i+1}^{n},Z^{i-1},K)$, and $(d)$ follows by applying the data processing inequality to the Markov chain in (\ref{eq:intermMarkov}).
		
\textbf{Privacy Leakage Rate}: We obtain
\begin{align}
&n(R_{\ell}+\delta_n)\nonumber\\ 
&\overset{(a)}{\geq}[I(W;Y^n)-I(W;Z^n)]+[I(W;X^n)-I(W;Y^n)]\nonumber\\
&\overset{(b )}{=}[I(W;Y^n)-I(W;Z^n)]\nonumber\\
&\qquad +I(W;X^n|K)-I(K;X^n|W)\nonumber\\
&\qquad-I(W;Y^n|K)+I(K;Y^n|W)\nonumber\\
&\overset{(c)}{=}[I(W;Y^n)-I(W;Z^n)]\nonumber\\
&\qquad +[I(W;X^n|K)-I(W;Y^n|K)]-I(K;X^n|W,Y^n)\nonumber\\
&\geq \sum_{i=1}^n\Big[I(W;Y_i|Y_{i+1}^n)-I(W;Z_i|Z^{i-1})\Big]\nonumber\\
&\qquad +\! \sum_{i=1}^n\!\Big[I(W;X_i|X^{i-1},K)\!-\!I(W;Y_i|Y_{i+1}^n,K)\Big]\!-\!H(K)\nonumber\\
&\overset{(d)}{=}\sum_{i=1}^n\Big[I(W;Y_i|Y_{i+1}^n,Z^{i-1})-I(W;Z_i|Z^{i-1},Y_{i+1}^n)-R_0\Big]\nonumber\\
&\qquad+ \sum_{i=1}^n\Big[I(W;X_i|X^{i-1},Y_{i+1}^n,K)\nonumber\\
&\qquad\qquad\qquad-I(W;Y_i|Y_{i+1}^n,X^{i-1},K)\Big]\nonumber\\
&\overset{(e)}{=}\!\sum_{i=1}^n\Big[I(W;Y_i|Y_{i+1}^n,Z^{i-1})\!-\!I(W;Z_i|Z^{i-1},Y_{i+1}^n)\!-\!R_0\Big]\nonumber\\
&\qquad+ \sum_{i=1}^n\Bigg[I(W;X_i|X^{i-1},Y_{i+1}^n,Z^{i-1},K)\nonumber\\
&\qquad\qquad\qquad\qquad-I(W;Y_i|Y_{i+1}^n,X^{i-1},Z^{i-1},K)\Bigg]\nonumber\\
&\overset{(f)}{=}\sum_{i=1}^n\Big[I(W,Y_{i+1}^n,Z^{i-1};Y_i)\!-\!I(W,Z^{i-1},Y_{i+1}^n;Z_i)-R_0\Big]\nonumber\\
&\qquad+ \sum_{i=1}^n\Bigg[I(W,X^{i-1},Y_{i+1}^n,Z^{i-1},K;X_i)\nonumber\\
&\qquad\qquad\qquad\qquad-I(W,Y_{i+1}^n,X^{i-1},Z^{i-1},K;Y_i)\Bigg]\nonumber\\
&\overset{(g)}{=} \sum_{i=1}^n \Big[I(V_i;Y_i)-I(V_i;Z_i)-R_0\nonumber\\
&\qquad\qquad+I(U_i,V_i;X_i)-I(U_i,V_i;Y_i)\Big]\nonumber\\
&= \sum_{i=1}^n \Bigg[-I(U_i,V_i;Z_i)-R_0+I(U_i,V_i;X_i)\nonumber\\
&\qquad\qquad+\left(I(U_i;Z_i|V_i)-I(U_i;Y_i|V_i)\right)\Bigg]\nonumber\\
&\overset{(h)}{\geq}\! \sum_{i=1}^n\Big[I(U_i;X_i|Z_i)-R_0\nonumber\\
&\qquad\qquad+\![I(U_i;Z_i|V_i)\!-\!I(U_i;Y_i|V_i)]^-\Big]
\end{align}
where $(a)$ follows by (\ref{eq:privleakagelossySC_cons}) and from the Markov chain $W-X^n-Z^n$, $(b)$ follows since $K$ is independent of $(X^n,Y^n)$, $(c)$ follows from the Markov chain $(W,K)-X^n-Y^n$, $(d)$ follows because $H(K)=nR_0$ and from Csisz\'{a}r's sum identity \cite{CsiszarKornerbook2011}, $(e)$ follows from the Markov chains
\begin{align}
	&Z^{i-1}- (X^{i-1},Y_{i+1}^n,K)-(X_i,W)\label{eq:xizi-1Markov}\\
	&Z^{i-1}- (X^{i-1},Y_{i+1}^n,K)-(Y_i,W)\label{eq:yizi-1Markov}
\end{align}
$(f)$ follows because $(X^n,Y^n,Z^n)$ are i.i.d. and $K$ is independent of $(X^n,Y^n,Z^n)$, $(g)$ follows from the definitions of $V_i$ and $U_i$, and $(h)$ follows from the Markov chain $V_i-U_i-X_i-Z_i$.
		
Next, we provide the matching converse for the privacy leakage rate in (\ref{eq:AchprivWbarbar}), which is achieved when $R_0\geq I(U;\widetilde{X}|Y,V)$. We have
\begin{align}
&n(R_{\ell}+\delta_n)\overset{(a)}{\geq}H(X^n|Z^n)-H(X^n|Z^n,W)\nonumber\\
&\overset{(b)}{=}H(X^n|Z^n)-\sum_{i=1}^{n}H(X_i|Z_i,Z^{i-1},X_{i+1}^n,W,Y_{i+1}^n)\nonumber\\
&\overset{(c)}{=} H(X^n|Z^n)-\sum_{i=1}^n H(X_i|Z_i,V_i,X^n_{i+1})\nonumber\\
&\overset{(d)}{\geq} \sum_{i=1}^{n}[H(X_i|Z_i)-H(X_i|Z_i,V_i)]\nonumber\\
&=\sum_{i=1}^{n}I(V_i;X_i|Z_i)
\end{align}
where $(a)$ follows by (\ref{eq:privleakagelossySC_cons}), $(b)$ follows from the Markov chain
\begin{align}
	(Z_{i+1}^n,Y_{i+1}^n)-(X_{i+1}^n,W,Z^i)-X_i
\end{align}
$(c)$ follows from the definition of $V_i$, and $(d)$ follows because $(X^n,Z^n)$ are i.i.d.

We remark that the matching converse for the privacy leakage rate in (\ref{eq:Achprivacyleakzero}), achieved when $R_0\geq I(U;\widetilde{X}|Y)$, follows from the fact that conditional mutual information is non-negative.

\textbf{Secrecy Leakage Rate}: We have
\begin{align}
&n(R_{\text{s}}+\delta_n)\nonumber\\ 
&\overset{(a)}{\geq}[I(W;Y^n)-I(W;Z^n)]+[I(W;\widetilde{X}^n)-I(W;Y^n)]\nonumber\\
&\overset{(b )}{=}[I(W;Y^n)-I(W;Z^n)]\nonumber\\
&\qquad +I(W;\widetilde{X}^n|K)-I(K;\widetilde{X}^n|W)\nonumber\\
&\qquad-I(W;Y^n|K)+I(K;Y^n|W)\nonumber\\
&\overset{(c)}{=}[I(W;Y^n)-I(W;Z^n)]\nonumber\\
&\qquad +[I(W;\widetilde{X}^n|K)-I(W;Y^n|K)]-I(K;\widetilde{X}^n|W,Y^n)\nonumber\\
&\overset{(d)}{\geq} \sum_{i=1}^n\Big[I(W;Y_i|Y_{i+1}^n)-I(W;Z_i|Z^{i-1})\Big]\nonumber\\
&\qquad +I(W;\widetilde{X}^n|Y^n,K)-\!H(K)\nonumber
\end{align}
\begin{align}
&\overset{(e)}{=}\sum_{i=1}^n\Big[I(W;Y_i|Y_{i+1}^n,Z^{i-1})-I(W;Z_i|Z^{i-1},Y_{i+1}^n)-R_0\Big]\nonumber\\
&\qquad+ nH(\widetilde{X}|Y)- \sum_{i=1}^{n}H(\widetilde{X}_i|Y_i,Y^n_{i+1},W,K,\widetilde{X}^{i-1})\nonumber\\
&\overset{(f)}{\geq}\sum_{i=1}^n\Big[I(W,Y_{i+1}^n,Z^{i-1};Y_i)-I(W,Z^{i-1},Y_{i+1}^n;Z_i)-R_0\Big]\nonumber\\
&\qquad+ nH(\widetilde{X}|Y)- \sum_{i=1}^{n}H(\widetilde{X}_i|Y_i,Y^n_{i+1},W,K,X^{i-1},Z^{i-1})\nonumber\\
&\overset{(g)}{=}\sum_{i=1}^n\Big[I(V_i;Y_i)-I(V_i;Z_i)-R_0\Big]\nonumber\\
&\qquad+ nH(\widetilde{X}|Y)- \sum_{i=1}^{n}H(\widetilde{X}_i|Y_i,U_i,V_i)\nonumber\\
&\overset{(h)}{=}\sum_{i=1}^n\Big[I(V_i;Y_i)-I(V_i;Z_i)-R_0\Big]\nonumber\\
&\qquad+\sum_{i=1}^{n}\Big[I(U_i,V_i;\widetilde{X}_i)-I(U_i,V_i;Y_i)\Big]\nonumber\\
&= \sum_{i=1}^n \Bigg[-I(U_i,V_i;Z_i)-R_0+I(U_i,V_i;\widetilde{X}_i)\nonumber\\
&\qquad\qquad+\left(I(U_i;Z_i|V_i)-I(U_i;Y_i|V_i)\right)\Bigg]\nonumber\\
&\overset{(i)}{\geq}\! \sum_{i=1}^n\Big[I(U_i;\widetilde{X}_i|Z_i)-R_0\nonumber\\
&\qquad\qquad+\![I(U_i;Z_i|V_i)\!-\!I(U_i;Y_i|V_i)]^-\Big]
\end{align}
where $(a)$ follows by (\ref{eq:secrecyleakagelossySC_cons}) and from the Markov chain $W-\widetilde{X}^n-Z^n$, $(b)$ follows because $K$ is independent of $(\widetilde{X}^n,Y^n)$, $(c)$ and $(d)$ follow from the Markov chain $(W,K)-\widetilde{X}^n-Y^n$, $(e)$ follows because $H(K)=nR_0$ and $(\widetilde{X}^n,Y^n)$ are i.i.d. and independent of $K$, and from the Csisz\'{a}r's sum identity and the Markov chain
\begin{align}
	Y^{i-1}-(\widetilde{X}^{i-1},W,K,Y_{i+1}^n,Y_i)-\widetilde{X}_{i}
\end{align}
$(f)$ follows since $(Y^n,Z^n)$ are i.i.d. and from the data processing inequality applied to the Markov chain 
\begin{align}
(X^{i-1},Z^{i-1})-(\widetilde{X}^{i-1},W,K,Y_{i+1}^n,Y_i)-\widetilde{X}_{i}
\end{align}
$(g)$ follows from the definitions of $V_{i}$ and $U_{i}$, $(h)$ follows from the Markov chain $(V_i,U_i)-\widetilde{X}_i-Y_i$, and $(i)$ follows from the Markov chain $V_i-U_i-\widetilde{X}_i-Z_i$.

Next, the matching converse for the secrecy leakage rate in (\ref{eq:AchsecrecyWbarbar}), achieved when $R_0\geq I(U;\widetilde{X}|Y,V)$, is provided.
\begin{align}
&n(R_{\text{s}}+\delta_n)\overset{(a)}{\geq}H(\widetilde{X}^n|Z^n)-H(\widetilde{X}^n|Z^n,W)\nonumber\\
&\overset{(b)}{\geq}H(\widetilde{X}^n|Z^n)-\sum_{i=1}^{n}H(\widetilde{X}_i|Z_i,Z^{i-1},\widetilde{X}_{i+1}^n,W,Y_{i+1}^n)\nonumber\\
&\overset{(c)}{=} H(\widetilde{X}^n|Z^n)-\sum_{i=1}^n H(\widetilde{X}_i|Z_i,V_i,\widetilde{X}^n_{i+1})\nonumber
\end{align}
\begin{align}
&\overset{(d)}{\geq} \sum_{i=1}^{n}[H(\widetilde{X}_i|Z_i)-H(\widetilde{X}_i|Z_i,V_i)]=\sum_{i=1}^{n}I(V_i;\widetilde{X}_i|Z_i)
\end{align}
where $(a)$ follows by (\ref{eq:secrecyleakagelossySC_cons}), $(b)$ follows from the Markov chain
\begin{align}
(Z_{i+1}^n,Y_{i+1}^n)-(\widetilde{X}_{i+1}^n,W,Z^i)-\widetilde{X}_i
\end{align}
$(c)$ follows from the definition of $V_i$, and $(d)$ follows because $(\widetilde{X}^n,Z^n)$ are i.i.d.

Similar to the privacy leakage analysis above, the matching converse for the secrecy leakage rate in (\ref{eq:Achsecrecyleakzero}), achieved when $R_0\geq I(U;\widetilde{X}|Y)$, follows from the fact that conditional mutual information is non-negative.
\end{IEEEproof}

Introduce a uniformly distributed time-sharing random variable $\displaystyle Q\!\sim\! \text{Unif}[1\!:\!n]$ that is independent of other random variables, and define $X\!=\!X_Q$, $\displaystyle \widetilde{X}\!=\!\widetilde{X}_{Q}$, $\displaystyle Y\!=\!Y_Q$, $\displaystyle Z\!=\!Z_Q$, $V\!=\!V_{Q}$, and $U\!=\!(U_{Q},\!Q)$, so
\begin{align}
& (Q,V)\!-U-\widetilde{X}-X-(Y,Z)
\end{align}
form a Markov chain. The converse proof follows by letting $\delta_n\rightarrow0$.

\textbf{Cardinality Bounds}: We use the support lemma \cite[Lemma 15.4]{CsiszarKornerbook2011} for the cardinality bound proofs, which is a standard step, so we omit the proof.

\section*{Acknowledgment}
O. G{\"u}nl{\"u} and R. F. Schaefer were supported in part by the German Federal Ministry of Education and Research (BMBF) under the Grant 16KIS1004. H. Boche was supported in part by the BMBF within the national initiative on 6G Communication Systems through the research hub 6G-life under the Grant 16KISK002 and within the national initiative on Information Theory for Post Quantum Crypto ``Quantum Token Theory and Applications - QTOK" under the Grant 16KISQ037K, which has received additional funding from the German Research Foundation (DFG) within Germany’s Excellence Strategy EXC-2092 CASA-390781972. H. V. Poor was supported in part by the U.S. National Science Foundation (NSF) under the Grant CCF-1908308.

\IEEEtriggeratref{41}
\bibliographystyle{IEEEtran}
\bibliography{references_ITW2022}

\end{document}